\documentclass[11pt,twoside]{article}  
\usepackage{asp2006}
\usepackage{adassconf}

\usepackage{latexsym}
\usepackage{amsthm}
\usepackage{amsmath}
\usepackage{amssymb}
\usepackage{amsfonts}

\begin{document}   

%

\paperID{P5.10}

%

\title{Photometric Redshift Estimation on SDSS Data Using Random Forests}
       
%
%
%
%
%

\markboth{Carliles, Budav\'ari, Heinis, Priebe, and Szalay}{Photometric Redshift Estimation on SDSS Data Using Random Forests}

%

\author{Samuel\ Carliles\altaffilmark{1}, Tam\'as\ Budav\'ari\altaffilmark{2},
              Sebastien\ Heinis\altaffilmark{2}, Carey\ Priebe\altaffilmark{3},
              Alexander\ Szalay\altaffilmark{2}}


\altaffiltext{1}{Dept. of Computer Science, Johns Hopkins University, Baltimore, MD, USA}
\altaffiltext{2}{Dept. of Physics and Astronomy, Johns Hopkins University, Baltimore, MD, USA}
\altaffiltext{3}{Dept. of Applied Mathematics and Statistics, Johns Hopkins University, Baltimore, MD, USA}


\contact{Samuel Carliles}
\email{ihsahn@cs.jhu.edu}

%
%

\paindex{Carliles, S.}
\aindex{Budav\'ari, T.}
\aindex{Heinis, S.}
\aindex{Priebe, C.}
\aindex{Szalay, A.}






\begin{abstract}          
Given multiband photometric data from the SDSS DR6, we estimate galaxy redshifts.
We employ a Random Forest trained on color features and spectroscopic redshifts from
80,000 randomly chosen primary galaxies yielding a mapping from color to redshift such
that the difference between the estimate and the spectroscopic redshift is small.  Our
methodology results in tight RMS scatter in the estimates limited by photometric errors.
Additionally, this approach yields an error distribution that is nearly Gaussian with
parameter estimates giving reliable confidence intervals unique to each galaxy
photometric redshift.
\end{abstract}


\section{The Problem}
We are given five bands of photometric data from the Sloan Digital Sky Survey 
Data Release 6 \htmladdnormallinkfoot{(SDSS DR6)}{http://www.sdss.org/dr6/}.
Associated with each magnitude measurement is
an error measurement and an extinction measurement used to correct the effect
of Galactic dust.  For some objects we have
spectroscopic redshifts, hereon denoted $z_{spec}$ for a particular object.  We wish
to estimate photometric redshifts for non-spectroscopic objects which are represented
by the spectroscopic sample.  We are currently interested only
in galaxies with $10^{-4} \le z_{spec} \le 1$, and we
exclude other objects which are likely to "contaminate" our sample.  We further choose
for the moment not to consider Luminous Red Galaxies (LRGs), leaving us with some
$400,000$ objects with which to train and test.  In our current work we use
magnitudes available in the new Ubercal table (Padmanabhan 2007) for each object.  We subtract
sequential extinction-corrected magnitudes to get color features, named $u$, $g$, $r$, and $i$.
For instance, our color $u$ is actually the magnitude difference $u - g$ and so on.

\section{Our Proposed Solution}
Given photometric colors and spectroscopic redshifts
$(u_j, g_j, r_j, i_j, z_{spec_j}) \in \mathbb{R}^5$,
$j \in \{1, 2, \ldots, n_{spec}\}$ where $n_{spec}$ is the number of objects for which
we have spectroscopic redshifts, and assuming the existence of a
function $f : \mathbb{R}^4 \rightarrow \mathbb{R}$ mapping tuples
$(u_j, g_j, r_j, i_j)$ to unique $z_{spec_j}$ values, estimating $f$ becomes a regression
problem, and we can choose from among any number of common methods of
regression.  We choose Random Forest regression.

\textbf{Random Forests} (Breiman 2001) are ensembles of Classification and Regression
Trees (Breiman 1984) trained on bootstrap samples (Efron 1994).  A random forest is composed
of one or more regression trees.  Trees are grown
\textit{mostly} ``the normal way'', at each node making a binary partition at the ``best''
point along the ``best'' axis.  In random forests, however, rather than choosing the best
axis at each node from the input space, it is chosen from a random subspace of the
input space (Ho 1998).  Nodes are split until a user-specified minimum number of
inputs (commonly five) is reached in a node and that node is declared terminal, its
value being defined as the mean of the values of its constituent inputs in the regression
case.  After a tree is grown, a new input is classified left or right starting at the
root of the tree, moving down until it reaches its associated terminal node, and that node's
value becomes the tree estimate for that input.  The ensemble regression estimate for that
input is then generally taken to be the mean of the corresponding regression tree estimates.

\textbf{Our Procedure} is to select a training set of $80,000$ objects uniformly at random
and without replacement from our full sample.  With this we train a random forest of $600$
trees yielding for
each redshift to be estimated a set of individual tree estimates, called $\hat{z}_{tree_i}$,
$i \in \{1, 2, \ldots, 600\}$.  We have several options for aggregating these into a single
forest estimate, and we choose a trimmed mean computed as follows.  For a
given test input we compute $\hat{z}_{mean} \equiv \overline{\hat{z}_{tree}}$ and
the RMS estimated tree error (called $\hat{\sigma}_{mean}$) around $\hat{z}_{mean}$ over the
$\hat{z}_{tree_i}$, then compute the mean over those $\hat{z}_{tree_i}$ which are within
$2\hat{\sigma}_{mean}$ of $\hat{z}_{mean}$, giving $\hat{z}_{trim}$, which is our final
redshift estimate for that input.  Our primary measure of accuracy is then the RMS of
$\hat{z}_{trim} - z_{spec}$ over all inputs.

\begin{figure}[t]
\epsscale{0.75}
\plottwo{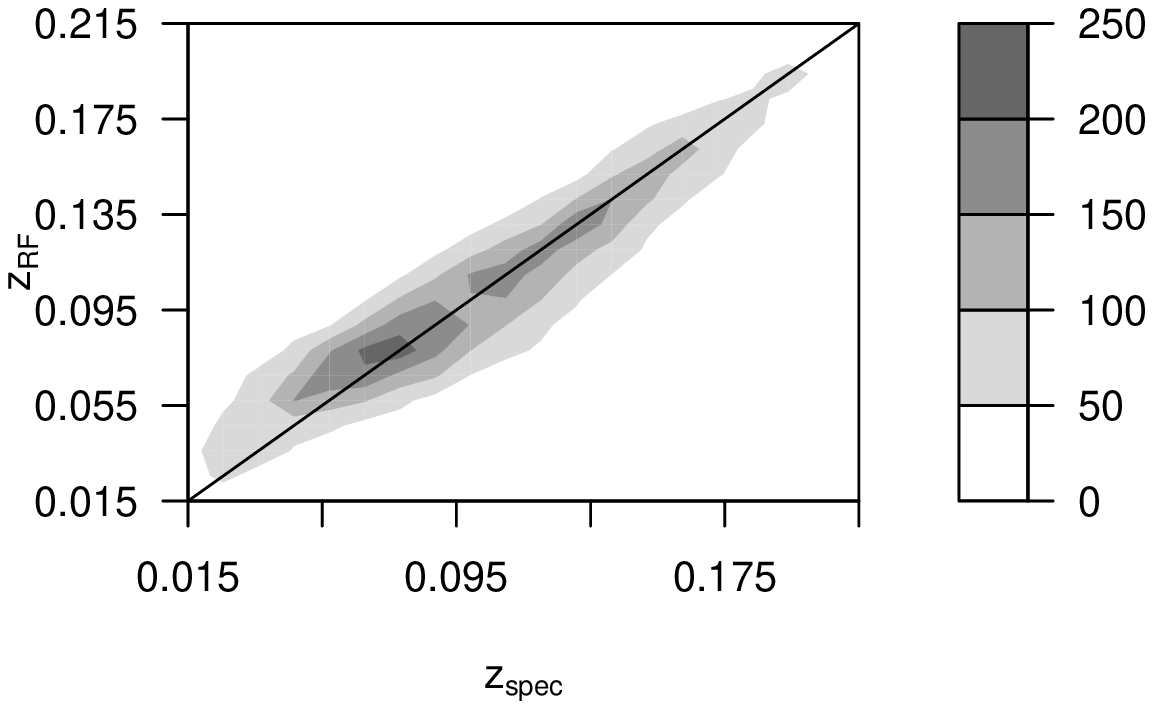}{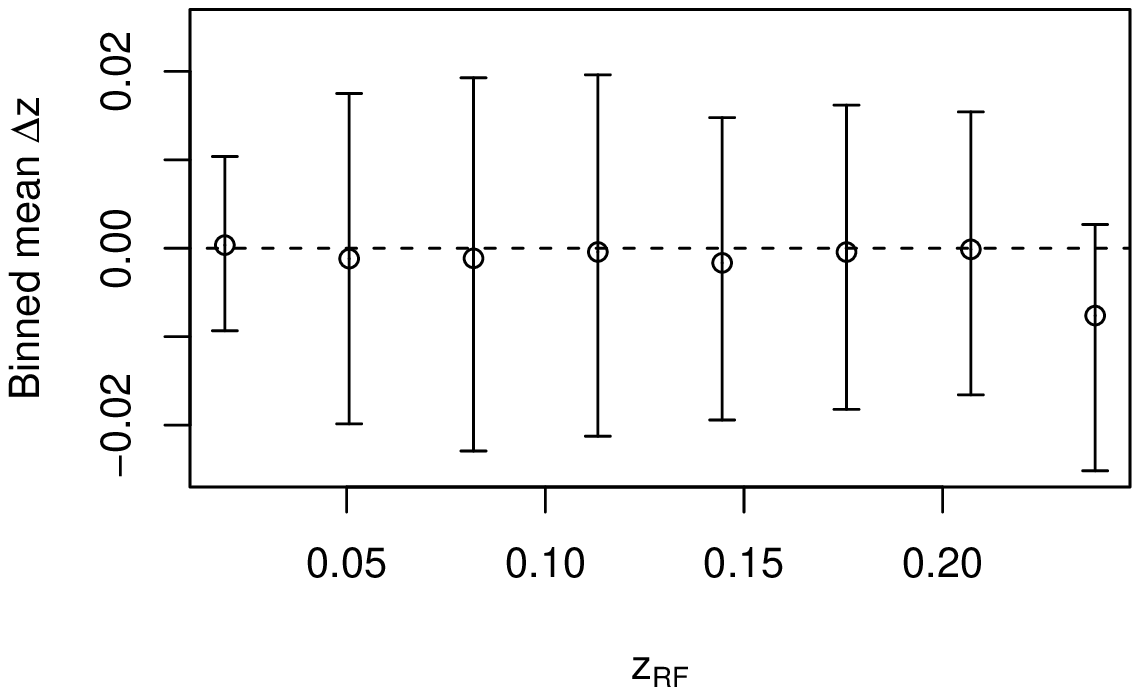}
\caption{a. (left) Photometric vs. spectroscopic redshift for 10,000 test objects. b. (right) Mean error for 10,000 test objects in eight bins vs. photometric redshift with bars marking region containing 34\% of errors on either side of mean.}\label{levelAndBars}
\end{figure}

\textbf{Our Error Model} falls out of the ensemble nature of random forests.  In the
following we use the untrimmed mean for notational simplicity, but the case with the trimmed
mean differs only trivially.  First define the regression tree errors for a given input as
$\epsilon_{tree_i} \equiv \hat{z}_{tree_i} - z_{spec}$, $i \in \{1, 2, \ldots, B\}$,
$B$ equal the number of trees in the forest ($600$ in our case).  Then define the random forest
regression error as $\epsilon_{mean} \equiv \hat{z}_{mean} - z_{spec}$, and it's trivial
to show that $\epsilon_{mean} = \overline{\epsilon_{tree}}$.  One can view the
tree errors as instances of identically distributed random variables.  Assuming
independence between trees (or at least a way to correct for dependence), one
can model the tree errors as random variables
$\varepsilon_{tree_i} \stackrel{\mathit{iid}}{\sim} \mathcal{F}(\mu, \sigma)$,
$i \in \{1, 2, \ldots, B\}$, with $\mathcal{F}$ some unknown distribution
with parameters $\mu$ and $\sigma$ \textit{unique to each input}.
Then the mean of these random variables is itself a random variable, call it
$\varepsilon_{mean}$, and the Central Limit Theorem gives us that
$\varepsilon_{mean} \sim N(\mu, \sigma / \sqrt{B})$,
thus we have a distribution for the random forest regression error for that input
and we have convenient plugin estimates for its parameters.  We then can further say that
\begin{eqnarray}
\frac{\varepsilon_{mean} - \mu}{\sigma / \sqrt{B}} \sim N(0, 1).
\end{eqnarray}
In practice we won't have $z_{spec}$ for most objects, however we do find that our
estimation error is zero mean (or, rather, insignificantly non-zero mean), so the best easy
estimate we have for $\mu$ is zero.  If, for a given input, we knew some better value for $\mu$,
it would mean our forest didn't utilize some information, and Figure~\ref{levelAndBars}b shows this not to be the case.  Our
estimate tells us nothing about whether we over- or under-estimated the actual value;
either outcome is equally likely conditioned on our redshift estimate.  But we do get
bias as a function of $z_{spec}$, and correcting this bias is a problem to be dealt with in future work.

\section{Our Results}
\begin{figure}[t]
\epsscale{0.75}
\plottwo{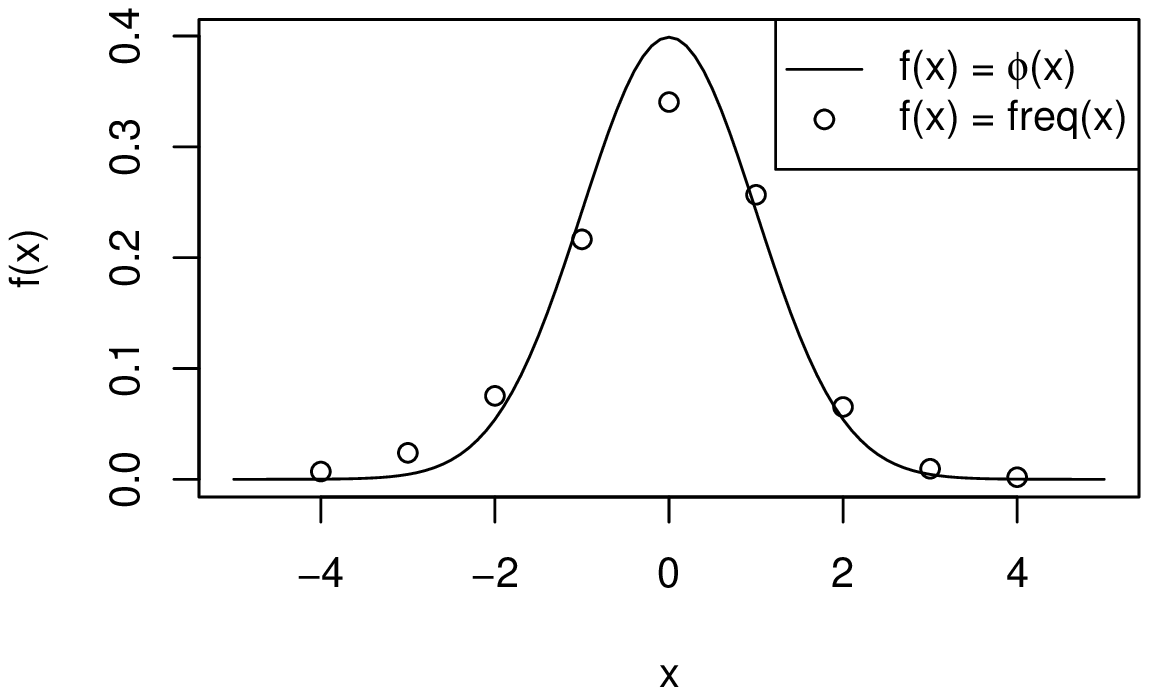}{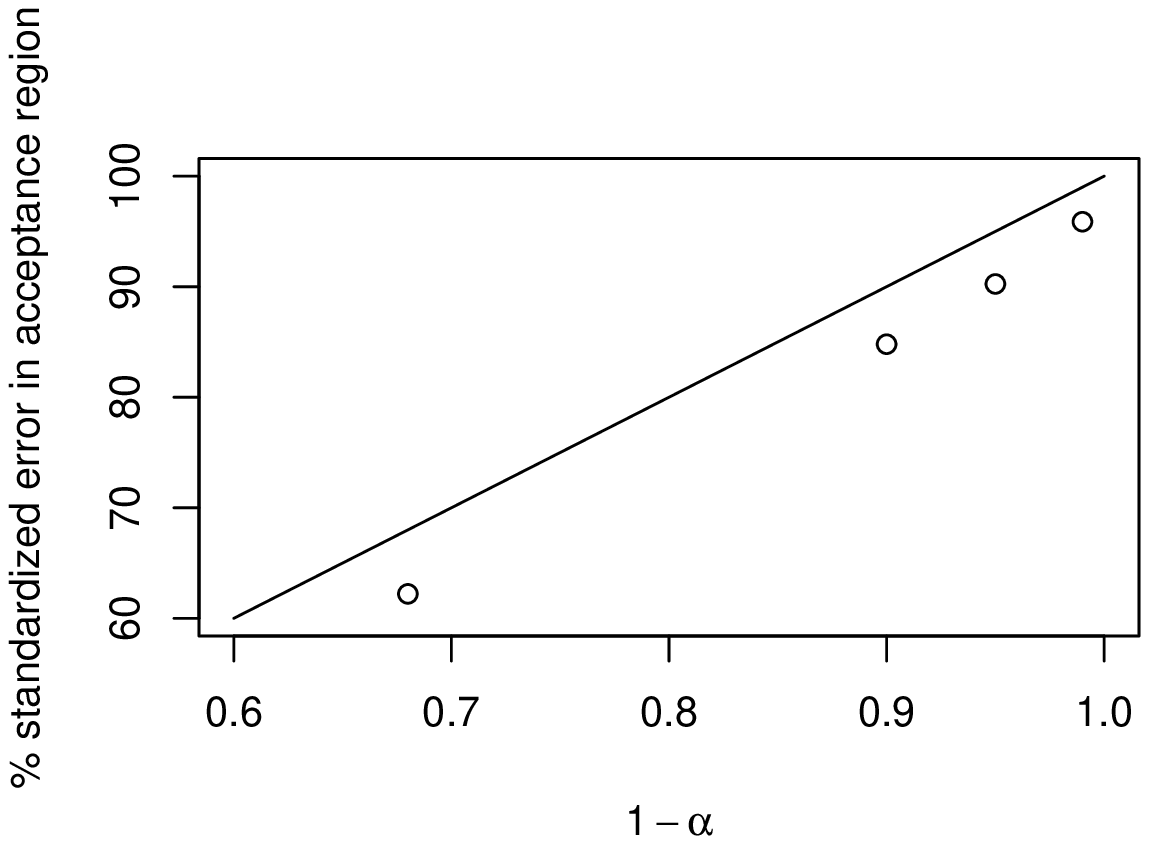}
\caption{a. (left) Observed standardized error for 10,000 test objects binned (circles) and Standard Normal distribution (curve). b. (right) Percent observed standardized error within level-$\alpha$ critical values for 10,000 test objects vs. $1 - \alpha$ (circles) and percent error expected within level-$\alpha$ critical values vs. $1 - \alpha$ (line).}\label{errDist}
\end{figure}

We select $10,000$ test objects uniformly at random from our held-out data (our main sample
minus our training set) and use our forest to produce redshift and error
distribution parameter
estimates for each test object.  The resulting RMS error between trimmed means
and corresponding $z_{spec}$ over these objects is $0.023$.  Figure~\ref{levelAndBars}a
characterizes our estimates, which are generally good with some slight bias visible near
the origin due to the local skewness of the underlying redshift distribution.
Figure~\ref{levelAndBars}b shows the binned mean difference between $z_{RF}$ and $z_{spec}$
as a function of $z_{RF}$ (where $z_{RF} \equiv \hat{z}_{trim}$ for each test input), and it indicates
that our random forest extracted nearly all the information contained in the training data.
Equation 1 above gives us a simple way to test our error distribution parameter estimates.
For each test input we set $z_{RF} = \hat{z}_{trim}$, $\epsilon_{mean} = z_{RF} - z_{spec}$
(pardon the flagrant abuse of notation),
$\mu = 0$, $\sigma = [\overline{\epsilon_{tree}^2}]^{1/2}$, and $B = 600$.  The results over
all test inputs should be distributed as iid
standard normal random variables.  However, in our tests, it turns out that setting $B = 600$,
the number of trees in our forest and the number of
ostensibly iid tree error estimates for each test input, yields a vertical spike after
standardizing.  In fact we discover empirically that in our case setting $B = 1$ yields
almost a perfect Standard Normal over all test inputs as shown in Figure~\ref{errDist}a.  We
interpret this to mean that we have strong dependence between tree error estimates, and
correcting for this algorithmically will be another subject of future work.  Moving on for now with the
parameter estimates resulting from our hand-chosen value of $B = 1$, we next seek to verify that
for given confidence levels $\alpha$, we have the expected number of errors between the
corresponding level-$\alpha$ critical values over all test inputs, and the quantile-quantile plot
shown in Figure~\ref{errDist}b is pleasing in that regard.  Thus we have good reason to believe
in our error distribution parameter estimates.

\section{Conclusion}
Our RMS error is consistent with results from previous studies using similar datasets, though slightly
lower RMS errors from different methodologies have been reported.  We may yet gain by expanding
our training data set as we have more than $300,000$ training objects in reserve and the computational
complexity of random forests is quite modest (in preliminary testing we trained multiple forests with
multiple training sets on a Core 2 Duo 2.0GHz MacBook in less than one week; regression on all of
our held-out data took several days using the same machine).  Future work will include better
accounting for dependence
between trees, investigating more deeply the behavior of our error distributions as a function of
$z_{spec}$, addressing $z_{spec}$-dependent bias, and extending estimation to objects not
represented by the training data.  As it is, the quality of the estimates, the per-object error distributions,
and the computational efficiency make this approach an attractive option for photometric redshift
estimation.

\end{document}